\documentclass[12pt]{article}

\usepackage{amssymb}
\usepackage{graphicx}
\usepackage{amsmath}
\usepackage{amsfonts,amsthm}

\usepackage[dvipdfmx]{color}

\newcommand{\argmin}{\mathop{\rm argmin}}

\usepackage{modnatbib}
\usepackage{url}

\usepackage{here}

\begin{document}

\title{Statistical learning for species distribution  models in ecological studies}

\author{Osamu Komori$^1$, Yusuke Saigusa$^2$ and Shinto Eguchi$^3$\\
$^1$Department of Computer and Information Science, Seikei University,Japan\\
$^2$Department of Biostatistics, School of Medicine, Yokohama City University, Japan\\
$^3$The Institute of Statistical Mathematics, Japan}

\maketitle

\begin{abstract}
We discuss species distribution models (SDM) for biodiversity studies in ecology. SDM plays an important role to estimate abundance of a species based on environmental variables that are closely related with the habitat of the species. The resultant habitat map indicates areas where the species is likely to live, hence it is essential for conservation planning and reserve selection. We especially focus on a Poisson point process and clarify relations with other statistical methods. Then we discuss a Poisson point process from a view point of information divergence, showing the Kullback-Leibler divergence of density functions reduces to the extended Kullback-Leibler divergence of intensity functions. This property enables us to extend the Poisson point process to that derived from other divergence such as $\beta$ and $\gamma$ divergences. Finally, we discuss integrated SDM and evaluate the estimating performance based on the Fisher information matrices.
\end{abstract}

\section{Introduction}
Interdisciplinary researches between statisticians, computer scientists and ecologists are important to develop new ideas and methodologies for  biodiversity studies. Their collaboration has produced many influential papers about species distribution modeling (SDM) \citep{Phillips2004}, hierarchical modeling and inference \citep{Royle2008}, species diversity measures \citep{Chao2005}, statistical modeling \citep{Thomas1991,Thomas2015} and so on. See the special feature of Methods in Ecology and Evolution \citep{Warton2015,Warton2017} for progress and achievements through collaboration of interdisciplinary research.

In this paper, we focus on SDM, especially on a Poisson point process, and review the related methods. Actually, a Poisson point process has close relationship with Maxent \citep{Phillips2004} and generalized linear model \citep{McCullagh1989}, where the estimated parameters are equivalent to each other except for an intercept term \citep{Renner2015}. This equivalence leads to an extension of a Poisson point process in which $L_1$ and $L_2$ penalties are applied in the estimation algorithm similar to the elastic net \citep{Fithian2013}. In fact, the estimating equation of a Poisson point process can be regarded as that of a weighted Poisson regression model as well as a weighted logistic regression model. And the weight functions lead to a connection to a weighted logistic regression model as well as an asymmetric logistic regression model \citep{Komori2016}. Moreover, we discuss the extension of a Poisson point process using quasi-linear modeling \citep{Komori2020} and $\beta$-divergence \citep{Basu1998, Minami2002}. Estimations of species distributions by a Poisson point process and other relating methods are illustrated using {\it Bradypus variegatus} data.

We also discuss a Poisson point process from a perspective of information divergence. We show that the Kullback-Leibler divergence between density functions reduces to the extended Kullback-Leibler divergence between intensity functions in a Poisson point process, ensuring the consistency of the estimator of a Poisson point process. This is established using an interesting property for random sum employed in the calculation of expectation in a Poisson point process \citep{Streit2010}. The relationship between density and intensity functions also gives rise to $\beta$-divergence between intensity functions, where a weight function in the estimating equation depends on the magnitude of the intensity function. Moreover, we extend the $\beta$-divergence to $U$-divergence \citep{Murata2004}, so that the parameters can be robustly estimated to outliers.

Then we discuss recent advances of SDM, called integrated SDM \citep{Koshkina2017,Fithian2015}. It combines presence-background data, sometimes referred to as presence-only data, and site-occupancy data, sometimes referred to as presence-absence data. The presence-background data is easily available from opportunistic surveys whereas it lacks in information on the absence of a species. On the other hand, site-occupancy data is of high quality because it possesses information on absence from planned surveys. We investigate the accuracy of the estimator of the integrated SDM, showing that simultaneous estimation of parameters of the integrated SDM is better than separate estimation of parameters of presence-background and site-occupancy models.

This paper is organized as follows. In the next section, we start with framework of a Poisson point process and review some methods closely related to a Poisson point process. Then we discuss a Poisson point process from a viewpoint of information divergence and the recent advances in section 3 and 4. We describe concluding remarks on a Poisson point process and biodiversity studies in the last section.
\section{Framework}

\subsection{Spatial Poisson point process}

We have a quick look at the framework for a Poisson point process, cf. \cite{Streit2010} for comprehensible introduction and practical applications.
Let $A$ be a subset of $\mathbb R^d$ to be provided observed points.
Then the event space is given by the collection of all possible finite subsets of $A$ as
\begin{eqnarray}
{\cal E}  =\{(0,\emptyset)\}\bigcup_{n=1}^\infty \{(n, \{s_1,...,s_n\}) \mid \{s_1,...,s_n\}\subset A)\},
\end{eqnarray}
where $\emptyset$ denotes an empty set.
Thus, the event space consists of pairs of the set of observed points $\{s_1,...,s_n\}$ and the number $n$.
Let $\lambda(s)$ be a positive function on $A$, called an intensity function.
A Poisson point process defined on ${\cal E} $ is described by the intensity function $\lambda(s)$ in a two-step procedure for any realization of ${\cal E} $. 
\begin{itemize}
\item[Step 1.] 
The number $n \geq0$ is determined by sampling the Poisson random variable, denoted by 
$N$, with probability mass function given by
\begin{eqnarray}\label{poisson}
p_N(n)=\frac{\Lambda ^n}{n!}\exp\{-\Lambda \}
\end{eqnarray}
where $\Lambda =\int_A \lambda(s)ds$ with an intensity function $\lambda(s)$ on $A$.
If $n = 0$, the realization is $\xi = (0, \emptyset)$, and Step 2 is not performed.

\item[Step 2.]
 For the $n$-point set $\{s_1,...,s_n\}$ the sequence $(s_1,...,s_n)$ is obtained by independent
and identically distributed  samples of a random variable $X$ on $A$ with
probability density function given by
\begin{eqnarray}\label{pi(s)}
\pi(s)=\frac{\lambda(s)}{\Lambda }
\end{eqnarray}
for $s \in A$. 

\end{itemize}

The procedure covers the basic statistical
structure of the Poison point process.  
It is noted that  in Step 2 the ordered pair $ (n, (s_1, . . . , s_n))$ is delicately different from the realization $\xi =(n, \{s_1, . . . , s_n\})$ such that $n!$ permuted vectors $(s_{\sigma(1)},...,s_{\sigma(n)})$'s are identified with the point set $\{s_1,...,s_n\}$, where $\sigma$ denotes a permutation on $\{1,...,n\}$.
For the joint random variable $\Xi=(N,\{X_1,...,X_N\})$, the density function is
written as
\begin{eqnarray}
p(\xi) &=& p_N(n) p_{\{X_1,...,X_N\}\mid N}(\{s_1,...,s_n\}\mid N=n)\nonumber\\
&=&
\frac{\Lambda ^n}{n!}\exp\{-\Lambda \}n!\prod_{i=1}^n \frac{\lambda(s_i)}{\Lambda }\nonumber\\
&=&
\exp\{-\Lambda \}\prod_{i=1}^n {\lambda(s_i)}.\label{f1}
\end{eqnarray}
The formula \eqref{f1} is surprisingly simple to introduce statistical procedures, in which 
the log-likelihood function is easily given in a tractable form.
Thus, the intensity function $\lambda(s)$ characterizes the distribution with the density function \eqref{f1} of the Poisson point process.
The set of all the intensity functions has a one-to-one correspondence with the set of all the distributions of the Poisson point processes. 
For example, we confirm that the intensity function is given by the density function as
\begin{eqnarray}
\lambda(s)=\frac{p(n+1,\{s,s_1,...,s_{n}\})}{p(n,\{s_1,...,s_n\})}
\end{eqnarray}
due to \eqref{f1}.

Let $N(A)$ be the number of points in a subset $A$ of $\mathbb R^d$. 
Then, $N(A)$ is distributed as the Poisson distribution in  \eqref{poisson} and $N(A_1),..., N(A_k)$ are independent
if $A_1,...,A_k$ are disjoint. 
It is known that this random function $N(\cdot)$ characterizes the Poisson point process that is defined by
Step 1 and Step 2.
We  will employ the notation $N(\cdot)$ in a subsequent discussion.

\subsection{Species distribution model}

Let us apply the framework of  Poisson point processes discussed above.
Assume that we get a presence dataset, say  $\{s_1,...,s_n\}$, or a set  of observed points for a species in a study area $A$.
Then, we build a statistical model of an intensity function that drives a Poisson point process on  ${A}$, in which a parametric model is given by 
\begin{eqnarray}\label{SDM}
{\cal M}=\{\lambda(  s, \theta): \theta\in\Theta\},
\end{eqnarray} 
called a species distribution model (SDM), where $ \theta$ is an unknown parameter in the space $\Theta$.
Typically, we shall consider a log -linear model $$\lambda(s,\theta)=\exp\{\theta_1^\top x(s)+\theta_0\}$$ with $\theta=(\theta_0,\theta_1)$,  a feature vector $x(s)$,  a slope vector $\theta_1$ and an intercept $\theta_0$.
Here $x(s)$ consists of geographical, climatic and other factors influencing the habitation of the species.  
Then, the log-likelihood function based on a realization  $(n,\{  s_1,...,  s_n\})$ is given by
\begin{eqnarray}\label{ell}
\ell( \theta)=
\sum_{i=1}^n  \log \lambda(  s_i,\theta)- \Lambda(\theta) 
\end{eqnarray}
due to \eqref{f1}, where $\Lambda( \theta)=\int_A\lambda(  s, \theta)d  s$.
Here the cumulative intensity is approximated as
\begin{eqnarray}
 \Lambda(\theta)
= \sum_{i=1}^m  w_i \lambda(  s_i, \theta)
\end{eqnarray}
by Gaussian quadrature, where ${  s_{n+1},...,   s_m}$ are the centers of the grid cells containing no presence location and $w_i$ is a quadrature weight for a grid cell area. Here $m$ denotes the total number of grid cells.
The approximate estimating equation is given by
\begin{eqnarray}\label{like}
{\cal E}(\theta) =
  \sum_{i=1}^m \{I(i\in\{1,\ldots,n\})-w_i \lambda(  s_i , \theta)\}
  \frac{\partial}{\partial \theta}\log  \lambda(  s_i , \theta) ={0},
\end{eqnarray}
where $I(\cdot)$ is the indicator function.
In a wide sense of SDM, the goal is to estimate the habitat probability of a species
across geographic space using the feature vectors $\{x(s):s\in A\}$.

\subsection{Statistical methods for SDMs}
\subsubsection*{Maxent}
In addition to a Poisson point process, Maxent \citep{Phillips2004} and logistic regression \citep{McCullagh1989} are also widely used for estimation of species distributions. Maxent models the species distribution $\pi(s)$ based on maximum entropy principle, where the entropy is defined as
\[
H_0(\pi)=-\sum_{i=1}^{m} \pi(s_i)\log\pi(s_i),
\]
where $\pi$ is defined in (\ref{pi(s)}) and the region $A$ is approximated by a finite set $\{s_1,...,s_m\}$.
The maximum entropy distribution based on environmental variables $x_1(s_i),\ldots,x_p(s_i)$ is derived from the following Lagrangian function with multipliers $\alpha_0,\alpha_{11},\ldots,\alpha_{1p}$:
\begin{equation}
\mathcal L(\pi)=H_0(\pi)+\sum_{j=1}^p\alpha_{1j}\Big\{\sum_{i=1}^m\pi(s_i)x_j(s_i)-\bar x_j\Big\}+\alpha_0\Big\{\sum_{i=1}^m\pi(s_i)-1\Big\},
\end{equation}
where $\bar x_j$ is the sample average of the environmental variable $x_j(s_i)$ over locations where the species is present:
\begin{equation}
\bar x_j=\frac1n\sum_{i=1}^n x_j(s_i),\ j=1,\ldots,p.
\end{equation}
Here we divided the total locations $s_1,\ldots,s_m$ into locations with presence $\{s_1,\ldots,s_n\}$ and others $\{s_{n+1},\ldots,s_m\}$ called background or pseudo-absence locations. Then we have

\begin{eqnarray*}
\frac{\partial\mathcal L(\pi)}{\partial\pi(s_i)}=-\log\pi(s_i)-1+\alpha_1^\top x(s_i)+\alpha_0=0,
\end{eqnarray*}
which leads to
\begin{equation}
\pi(s_i)=\exp(\alpha_1^\top x(s_i)+\alpha_0-1),
\end{equation}
where $\alpha_1^\top x(s_i)=\sum_{j=1}^p\alpha_{1j} x_j(s_i)$.
The term $\alpha_0-1$ corresponds to the standardization factor; hence, we have
\begin{equation}\label{pi_maxent}
\pi(s_i,\alpha_1)=\frac{\exp(\alpha_1^\top x(s_i))}{Z_{\alpha_1}},
\end{equation}
where $Z_{\alpha_1}=\sum_{i=1}^m\exp(\alpha_1^\top x(s_i))$, and we use a notation $\pi(x_i,\alpha_1)$ to clarify $\pi(x_i)$ is characterized by the parameter vector $\alpha_1=(\alpha_{11},\ldots,\alpha_{1p})^\top$ hereafter. In practice, $\alpha_1$ is estimated by the maximization of the log-likelihood:
\begin{equation}
\ell_{M}(\alpha_1)=\sum_{i=1}^n\log \pi(s_i,\alpha_1),
\end{equation}
resulting in the estimation equation regarding $\alpha_{1j}$ as
\begin{equation}\label{eq30}
\frac{\partial \ell_{M}(\alpha_1)}{\partial\alpha_{1j}}=\sum_{i=1}^mx_j(s_i)\bigg\{I(i\in\{1,\ldots,n\})-n\frac{\exp(\alpha_1^\top x(s_i))}{Z_{\alpha_1}}\bigg\}
\end{equation}
This indicates that $\alpha_1$ estimated by Maxent is equivalent to $\theta_1$ estimated by a Poisson point process because
\begin{equation}\label{eq31}
\frac{\partial \ell(\theta)}{\partial\theta_{1j}}=\sum_{i=1}^mx_j(s_i)\bigg\{I(i\in\{1,\ldots,n\})-w_i\exp(\theta_0+\theta_1^\top x(s_i))\bigg\},
\end{equation}
where $\ell(\theta)$ is the log-likelihood function by a Poisson point process defined in (\ref{ell}); $w_i$ is the quadrature weight for location $s_i$ and is replaced with $\lvert A \rvert/m$ which is the study area divided by sample size $m$. By comparing (\ref{eq30}) and (\ref{eq31}), we have
\begin{equation}
\alpha_1^\top x(s_i)+\log\frac{n}{Z_{\alpha_1}}=\theta_0+\theta_1^\top x(s_i)+\log \frac{\lvert A \rvert}{m},
\end{equation}
resulting in
\begin{equation}
\theta_{1j}=\alpha_{1j}\ (j=1,\ldots,p),\ and\ \theta_0=\log\frac{mn}{\lvert A \rvert Z_{\alpha_1}}.
\end{equation}
Note that $Z_{\alpha_1}$ depends on $\alpha_1$ but constant over $s_1,\ldots,s_n$. See \cite{Renner2013} for details of the proof.

If we put $L_1$-penalty to $\ell_{M}(\theta_1)$ to avoid overfitting, then the sequential algorithm for estimating $\theta_1$ is employed \citep{Dudik2004}. As for how to select the tuning parameter of $L_1$ penalty as well as functions of $x_j(s_i)$ such as linear, quadratic, threshold and hinge, see \cite{Phillips2008} for details.

We note that the estimating equation of (\ref{eq31}) can be regarded as that of weighted Poisson regression model because 
\begin{equation}
\frac{\partial \ell(\theta)}{\partial\theta_{1j}}=\sum_{i=1}^m w_i x_j(s_i)\bigg\{\zeta_i-\exp(\theta_0+\theta_1^\top x(s_i))\bigg\},
\end{equation}
where $\zeta_i=I(i\in\{1,\ldots,n\})/w_i$ and it can be regarded as a response variable. Hence the parameters $\theta$ in a Poisson point process can be estimated by iteratively reweighted least squares algorithm in the framework of generalized linear model \citep{McCullagh1989}.

\subsubsection*{Logistic regression model}
For a feature vector $x(s)=(x_1(s),\ldots,x_p(s))^\top$, a logistic regression model is formulated as
\begin{equation}
P(Y=1\mid x(s))=\frac{\exp(\beta_0+\beta_1^\top x(s))}{1+\exp(\beta_0+\beta_1^\top x(s))},
\end{equation}
where $Y$ is a random variable indicating presence of species $Y=1$ or absence $Y=0$. The probability $P(Y=1)$ is estimated by the number of presence locations divided by the total number of locations in the study area, that is $n/m$. If we consider $m\to\infty$ which correspond to $\beta_0\to-\infty$, then we have
\begin{equation}
P(Y=1\mid x(s))\approx \exp(\beta_0+\beta_1^\top x(s))\equiv p(s,\beta),
\end{equation}
where $\beta=(\beta_0,\beta_1)$.
In this setting, the log-likelihood of the logistic regression model is given as
\begin{eqnarray}
\ell_L(\beta)&=&\sum_{i=1}^m\log\{p(s_i,\beta)^{y_i}(1-p(s_i,\beta))^{1-y_i}\}\\&=&\sum_{i=1}^n\log p(s_i,\beta)+\sum_{n+1}^m\log(1-p(s_i,\beta))\\
&\approx&\sum_{i=1}^n\log p(s_i,\beta)-\sum_{n+1}^m p(s_i,\beta)\\
&\approx&\sum_{i=1}^m\Big\{I(i\in\{1,\ldots,n\})\log p(s_i,\beta)-p(s_i,\beta)\Big\},
\end{eqnarray}
resulting in the estimation equation
\begin{equation}
\frac{\partial \ell_{L}(\beta)}{\partial\beta_{1j}}=\sum_{i=1}^mx_j(s_i)\bigg\{I(i\in\{1,\ldots,n\})-\exp(\beta_0+\beta_1^\top x(s))\bigg\}.
\end{equation}
By comparing with (\ref{eq31}), when $m\to\infty$ we approximately have
\begin{equation}
\theta_{1j}=\beta_{1j}\ (j=1,\ldots,p),\ and\ \theta_0=\beta_0-\log \frac{\lvert A \rvert}{m}.
\end{equation}
See \cite{Warton2010} for details of the proof. A similar result is obtained by  \cite{Fithian2013}, considering a weighted logistic regression in which $\beta_0\to-\infty$ is implicitly assumed and infinite weights are employed to show the equivalence to the log-likelihood of a Poisson point process.

\subsubsection*{Weighted logistic regression model}
To deal with imbalance in sample sizes of classes $y=0$ and $y=1$ (in our case the number of locations of background $m-n$ and the number of those of presence $n$), weighted logistic regression model is recognized as useful \citep{Manski1977,King2001}, where weighed log-likelihood is used and the estimating equation is given as
\begin{equation}
\sum_{i=1}^m\omega_ix_j(s_i)\bigg\{y_i-\frac{\exp(\beta_0+\beta_1^\top x(s_i))}{1+\exp(\beta_0+\beta_1^\top x(s_i))}\bigg\},
\end{equation}
where $\omega_i$ is usually determined by the sample mean $\bar y$ and the population mean $\mu$. That is, $\omega_i=(\mu/\bar y)y_i+(1-\mu)/(1-\bar y)(1-y_i)$ for the observation $i$.
By applying under-sampling scheme to non-events, the value of $\bar y$ is usually set to around 0.5 in practice \citep{Maalouf2011,Maalouf2014}. However, this method is applicable only when the value of $\mu$ is known. However, the value of $\mu$ is unknown in general. In the case of infinitely weighted logistic regression \citep{Fithian2013}, $\omega_i$ is set to a large value such as $1000$ if $s_i$ is location of background $(y_i=0)$ and $0$ otherwise $(y_i=1)$. That is, it generates in a coercive manner a situation of imbalance in sample sizes $n\ll m$.

In contrast with the add hoc determination of $\omega_i$, $\omega_i$ is determined according to the linear predictor $\beta_0+\beta_1^\top x(s_i)$ in asymmetric logistic regression model \citep{Komori2016} as
\begin{equation}
\omega(s_i)=\frac{\exp(\beta_0+\beta_1^\top x(s_i))}{\exp(\beta_0+\beta_1^\top x(s_i))+\kappa},
\end{equation}
where $\kappa$ is a positive value. The weight is almost 1 when $\beta_0+\beta_1^\top x(s_i)$ takes a large positive value, which is the case for presence locations $y_i=1$; on the other hand, the weight goes to 0 when $\beta_0+\beta_1^\top x(s_i)$ takes a large negative value, which is the case for background locations. The weight in asymmetric logistic regression model is derived from the following conditional probability
\begin{equation}
P(Y=1\mid x(s))=\frac{\exp(\beta_0+\beta_1^\top x(s))+\kappa}{1+\exp(\beta_0+\beta_1^\top x(s))+\kappa},
\end{equation}
which corresponds to three-parameter logistic model in psychometrics \citep{Wainer2007}, and has a close relationship with a contamination model \citep{Copas1988}.

\subsubsection*{$\beta$-Maxent}
We are concerned with a restricted situation in which Maxent has a good performance to predict the habitation of a species.
In effect, the exponential model (\ref{pi_maxent}) is assumed as the maximum entropy distribution employing the classical entropy $H_0(\pi)$.
However, this model is not always correct to apply to the SDM.
So we consider the $\beta$-power entropy as
\begin{eqnarray}
  H_\beta(\pi)=-\frac{1}{\beta(\beta+1)} \sum_{i=1}^n \pi(s_i)^{\beta+1}.
\end{eqnarray} 
 Thus, the maximum entropy model derived from $H_\beta(\pi)$ is given as
\begin{equation}\label{beta-Gibbs}
\pi_\beta(s_i,\alpha_1)=\frac{\big\{1+\beta\alpha_1^\top x(s_i)\big\}^\frac1\beta}{Z_{\alpha_1}^\beta},
\end{equation}
by an argument similar to that with the classical entropy $H_0(\pi)$, where $Z_{\alpha_1}^\beta=\sum_{i=1}^m\{1+\beta\alpha_1^\top x(s_i)\}^{1/\beta}$ \citep{Komori2014,Komori2019,Eguchi2022}. We note that $H_0(\pi)$ is the limit of $H_\beta(\pi)$ as $\beta$ goes to 0. The model (\ref{pi_maxent}) is
called a deformed exponential model, cf. \citep{Naudts2011} for more broad
perspectives. The loss function derived from $\beta$-divergence \citep{Basu1998,Minami2002} is given as 
\begin{equation}
L_\beta(\alpha_1)=-\frac{1}{n\beta}\sum_{i=1}^n\{\pi_\beta(s_i,\alpha_1)^\beta-1\}+\frac1{1+\beta}\sum_{i=1}^m\pi_\beta(s_i,\alpha_1)^{1+\beta}.
\end{equation}
It is clear that $\lim_{\beta\to0}\pi_\beta(s_i,\alpha_1)=\pi(s_i,\alpha_1)$, which is the distribution of original Maxent defined in (\ref{pi_maxent}). A sequential algorithm for estimation of $\alpha_1$ is employed as in \cite{Komori2014}. The value of $\beta$ and the number of iteration of the sequential algorithm are determined by AIC for $M$-estimator \citep{Akaike1973,Konishi1996}, where the best $\beta$ is chosen in the range of $\{-1,-1/3,-1/5,0,1/5,1/3,1\}$ as in \cite{Eguchi2022}. Hence Kullback-Leibler divergence $(\beta=0)$ and the Itakura-Saito divergence ($\beta=-1$) are included in the analysis of $\beta$-Maxent.

\subsubsection*{Quasi-linear Poisson point process}
In a quasi-linear Poisson point process \citep{Komori2020}, the intensity function $\lambda(s,\theta,\alpha)$ is modeled based on Kolmogorov-Nagumo average \citep{Eguchi2015,Naudts2011} as
\begin{equation}
\lambda_\tau(s,\theta,\alpha)=\exp\bigg[\frac1\tau\log\bigg\{\frac12 \lambda(s,\theta)^\tau+\frac12 b(s,\alpha)^\tau\bigg\}\bigg],
\end{equation}
where $b(s,\alpha)$ denotes the sampling bias or imperfect detection modeled as $b(s,\alpha)=\exp(\alpha^\top z(s))$; $z(s)$ is a covariate vector relating to sampling bias such as distance to a road, to the nearest town or to the coast \citep{Fithian2015,Koshkina2017}. As a special case, it includes
\begin{equation}
\lim_{\tau \to 0}\lambda_\tau(s,\theta,\alpha)=\lambda(s,\theta)^\frac12b(s,\alpha)^\frac12,
\end{equation}
which is a thinned Poisson point process \citep{Fithian2015}. Also it includes
\begin{equation}
\lim_{\tau \to 1}\lambda_\tau(s,\theta,\alpha)=\frac12\lambda(s,\theta)+\frac12b(s,\alpha),
\end{equation}
which is a superposed Poisson point process \citep{Streit2010}. \cite{Komori2020} demonstrated a practical utility of a harmonic mean version formulated as
\begin{equation}
\lim_{\tau \to -1}\lambda_\tau(s,\theta,\alpha)=\frac{2}{\frac1{\lambda(s,\theta)}+\frac1{b(s,\alpha)}}
\end{equation}
using vascular plant data collected in Japan \citep{Kubota2015}. It has double weighted estimation equations
\begin{eqnarray}
\frac{\partial \ell_Q(\theta,\alpha)}{\partial \theta}
&=&\sum_{i=1}^mw_i\omega(s_i)x(s_i)\{\zeta_i-\lambda_\tau(s,\theta,\alpha)\}\\
\frac{\partial \ell_Q(\theta,\alpha)}{\partial \alpha}
&=&\sum_{i=1}^mw_i(1-\omega(s_i))z(s_i)\{\zeta_i-\lambda_\tau(s,\theta,\alpha)\},
\end{eqnarray} 
where $\omega(s_i)=\exp(\tau\theta^\top x(s_i))/\{\exp(\tau\theta^\top x(s_i))+\exp(\tau\alpha^\top z(s_i))\}$ and $\ell_Q(\theta,\alpha)=\sum_{i=1}^mw_i\{\zeta_i\log(\lambda_\tau(s,\theta,\alpha))-\lambda_\tau(s,\theta,\alpha)\}$. $1-\omega(s_i)$ shows a proportion of sampling bias effects at a location $s_i$. By plotting $1-\omega(s_i)$, $i=1,\ldots,m$ over a study region, we can identify areas heavily affected by sampling bias.

\subsection{Presence-only, presence-absence and abundance data}

SDM is motivated by a variety of objectives, including species conservation and management planning, monitoring for endangered and invasive species, and understanding species ecology, and so on \citep{Fithian2013}.
Species observation information is essential to predict the SDM.
Most data sets of species observation were obtained incidentally and therefore do not have reliable information about their absence.
Data without information about the absence of a species is called presence-only (PO) data.
PO data includes atlases, museum and herbarium records, incidental observation databases and radio-tracking studies, for example \citep{Elith2006}.
On the other hand, there are more reliable data sets that have information on the absence of a species through dedicated surveys conducted by surveyors with expertise.
For example, a surveyor spends one hour surveying a one-kilometer square area and records the presence or absence of a species.
Data with both presence and absence observations are called presence-absence (PA) data. 
In addition, data sets with information on the number of individuals are called abundance data (or count sampling data).
The development of geographic information systems has made it possible to obtain high-resolution environmental data, including geographic, climatic, and urbanization information necessary to predict species distributions.
The need for study on modeling methodologies using PO data increases since the PO data are now readily available without field surveys.
The PO data based on incidental discoveries often suffer from sampling bias due to human accessibility. 
Therefore, a number of methods have been investigated to address the sampling bias \citep{Dudik2005,Fithian2013,Komori2020}.
In addition, contamination of suspect data (outliers) due to incorrect geo-coordinates, taxonomic misclassification and shifts can be a problem, and screening tools are being developed, but manual work by experts is considered essential \citep{Belbin2013,Mesibov2013}. 
The problem is that time-consuming manual checking is prohibitively expensive in screening huge amounts of data, such as on a national or global scale.
Some methods discussed in this paper may be useful as a solution against the contamination of outliers.
In recent years, attempts have been made to improve species distribution predictions by combining data sets of different standards. 

\subsection{Estimation of species distribution using {\it Bradypus variegatus} data}
We demonstrate the estimation of species distribution by a Poisson point process and other relating methods using {\it Bradypus variegatus} data, which is the same data in \cite{Phillips2004} and is available {\tt dismo} package of statistical software R. The number of presence observation is $n=116$, and the number of grid cells is $m=5171$. We use 8 environmental variables from the WorldClim database such as mean annual temperature, total annual precipitation, precipitation of wettest quarter, precipitation of driest quarter, max temperature of warmest month, min temperature of coldest month, temperature annual range, and mean temperature of wettest quarter. In a quasi-linear Poisson point process, we only use an intercept term for bias modeling. The optimal tuning parameter $\beta$ in $\beta$-Maxent is selected among $\{-1,-1/3,-1/5,0,1/5,1/3,1\}$ based on AIC as in \citep{Eguchi2022}. AIC is also used for variable selection for all methods.

\begin{figure}[H]
  \begin{minipage}[b]{0.45\linewidth}
    \centering
    \includegraphics[width=7cm, height=8cm]{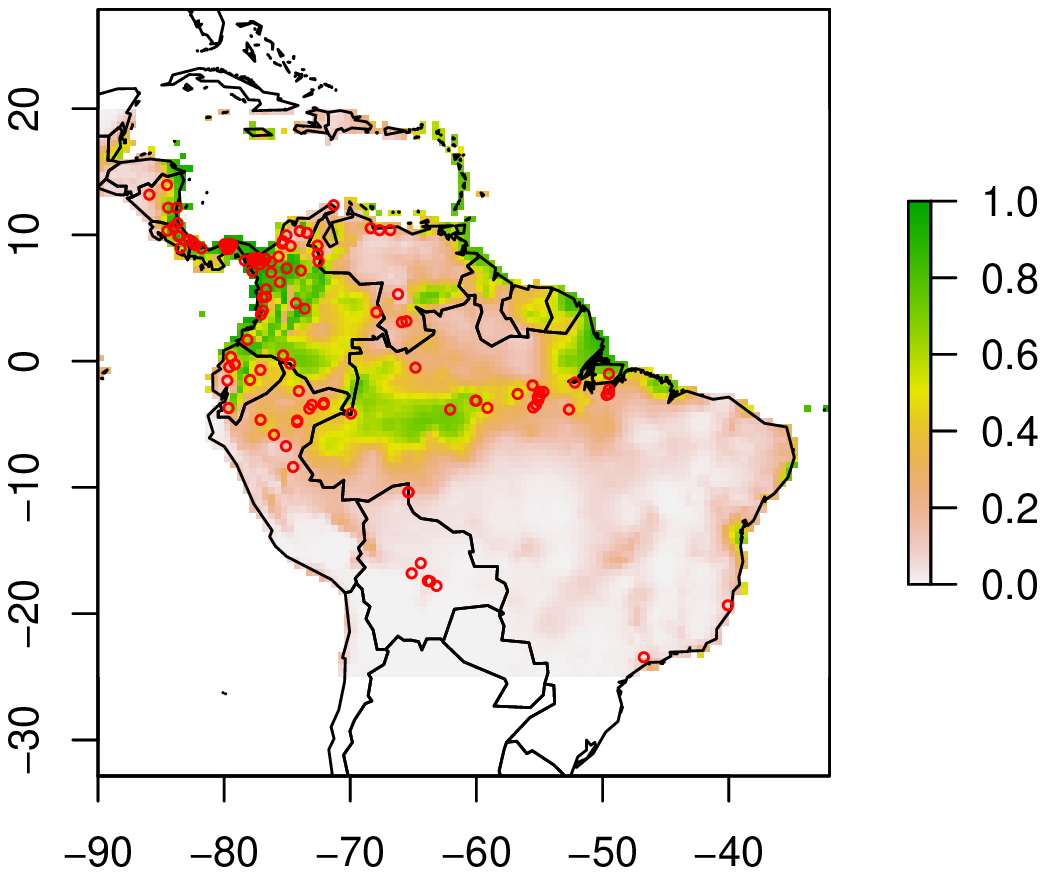}\\
 (a) Poisson point process (Maxent)\\
AUC=0.768,\ AIC=863.4
  \end{minipage}
  \begin{minipage}[b]{0.45\linewidth}
    \centering
    \includegraphics[width=7cm, height=8cm]{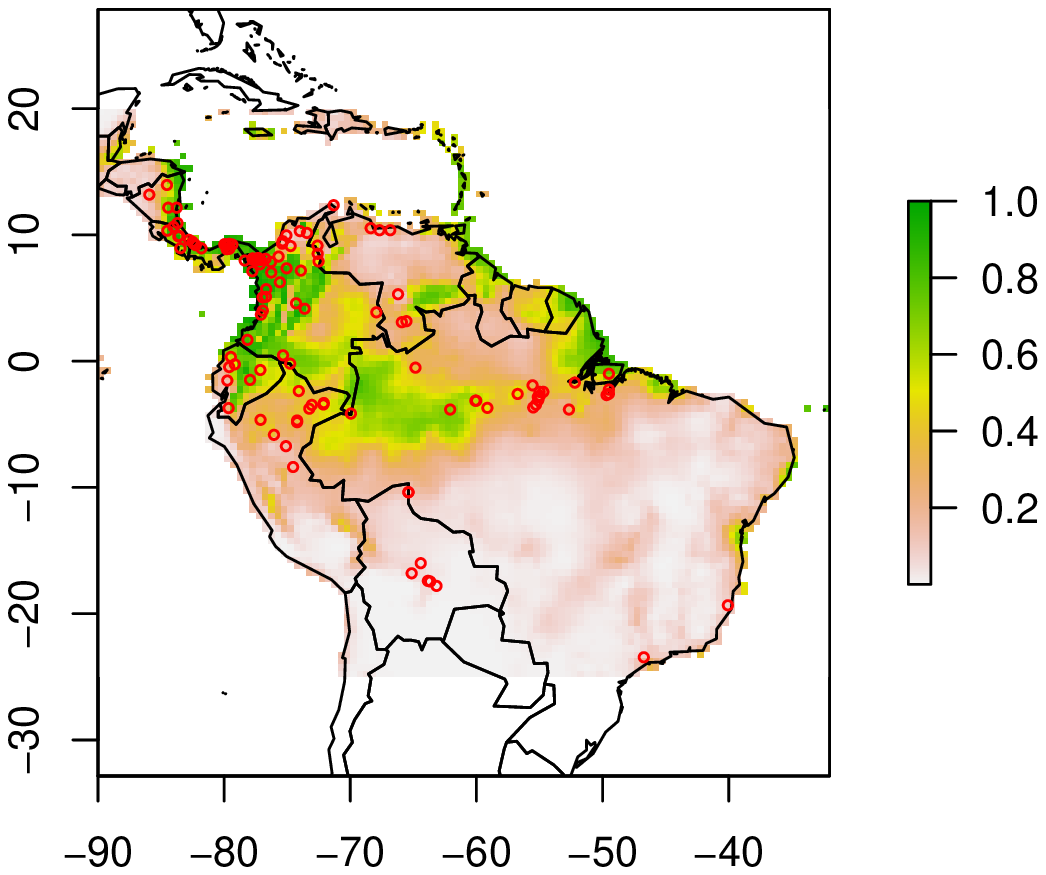}\\
 (b) Infinitely weighted logistic regression\\
AUC=0.778
  \end{minipage}

  \begin{minipage}[b]{0.45\linewidth}
    \centering
\includegraphics[width=7cm, height=8cm]{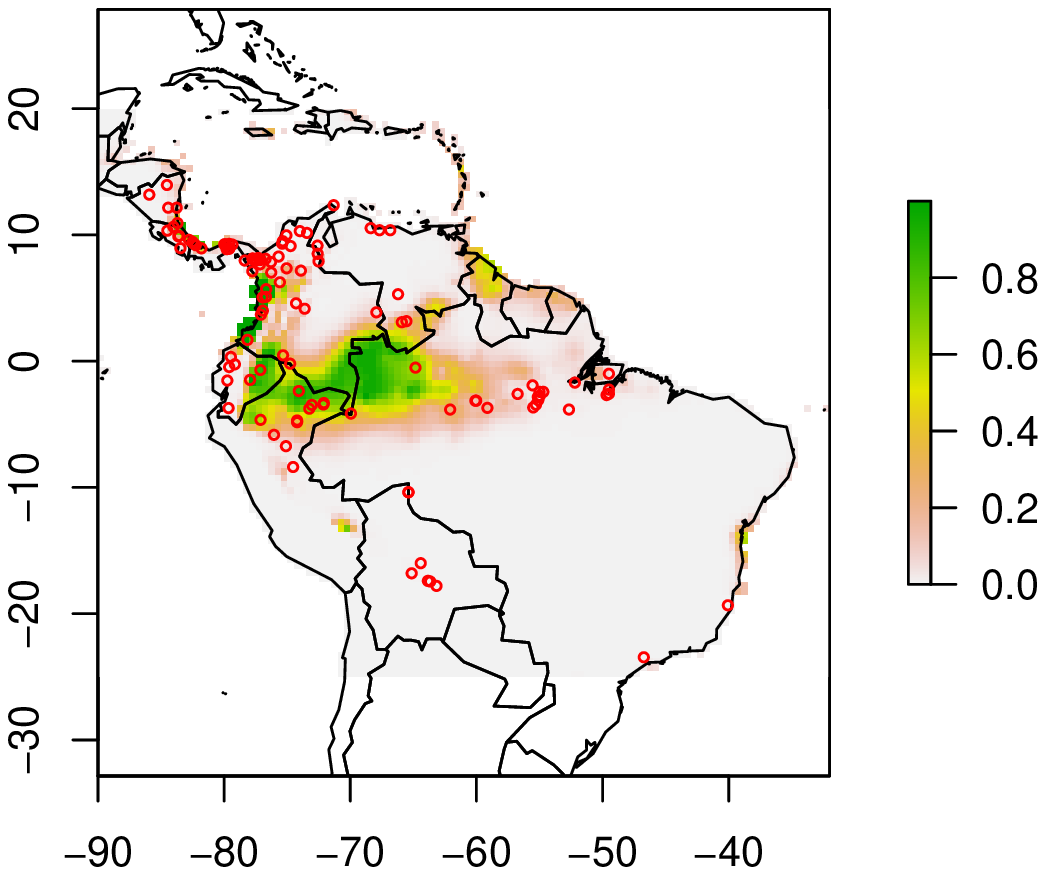}\\
 (c) Quasi-linear Poisson point process\\
AUC=0.761,\ AIC=856.9,\ $\hat\tau=-1$
  \end{minipage}
  \begin{minipage}[b]{0.45\linewidth}
    \centering
     \includegraphics[width=7cm, height=8cm]{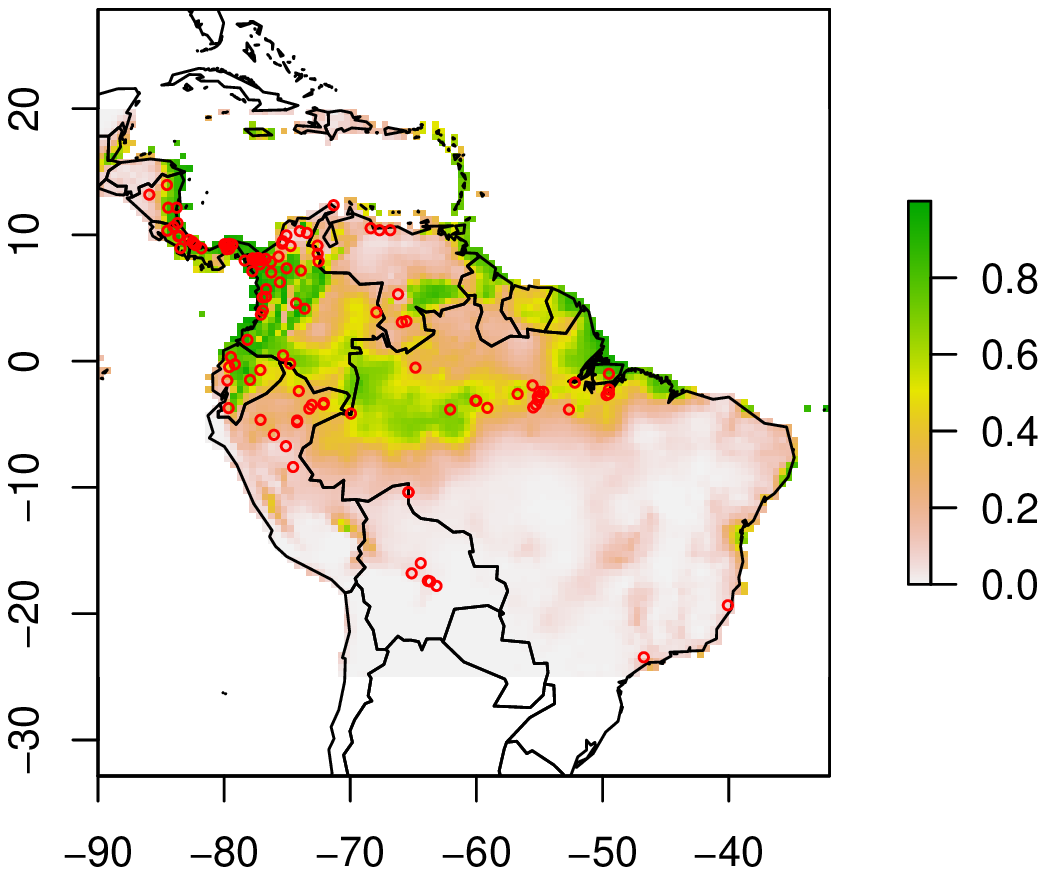}\\
 (d) $\beta$-Maxent\\
AUC=0.773, $\hat\beta=1/5$
  \end{minipage}

  \caption{Estimation of distribution of {\it Bradypus variegatus}}\label{fig1}
\end{figure}

Figure \ref{fig1} illustrates the estimation of species distribution for a Poisson point process, Infinitely weighted logistic regression, quasi-linear Poisson regression and $\beta$-Maxent. Observed locations of {\it Bradypus variegatus} are dotted in red. Clearly, areas with high estimated probabilities well correspond to the observed locations for all methods. As expected, the estimated distribution of the infinitely weighted logistic regression well resembles that of a Poisson point process. Also a similar estimation result is obtained by $\beta$-Maxent, where the optimal $\beta$ turns out to be $1/5$. On the other hand, the result of quasi-linear Poisson regression is quite different from others. Green areas with high estimated probabilities are observed in only northwestern regions. This tendency occurs because the optimal $\tau$ is estimated to be $-1$, resulting a harmonic mean of intensity functions. In fact, the model fitting of a quasi-linear Poisson point process is better than that of a Poisson point process, where the values of AIC based on $\ell_Q(\theta,\alpha)$ and $\ell(\theta)$ are 855.9 and 863.4, respectively. 

\begin{figure}[H]
\centering
    \includegraphics[width=15cm, height=10cm]{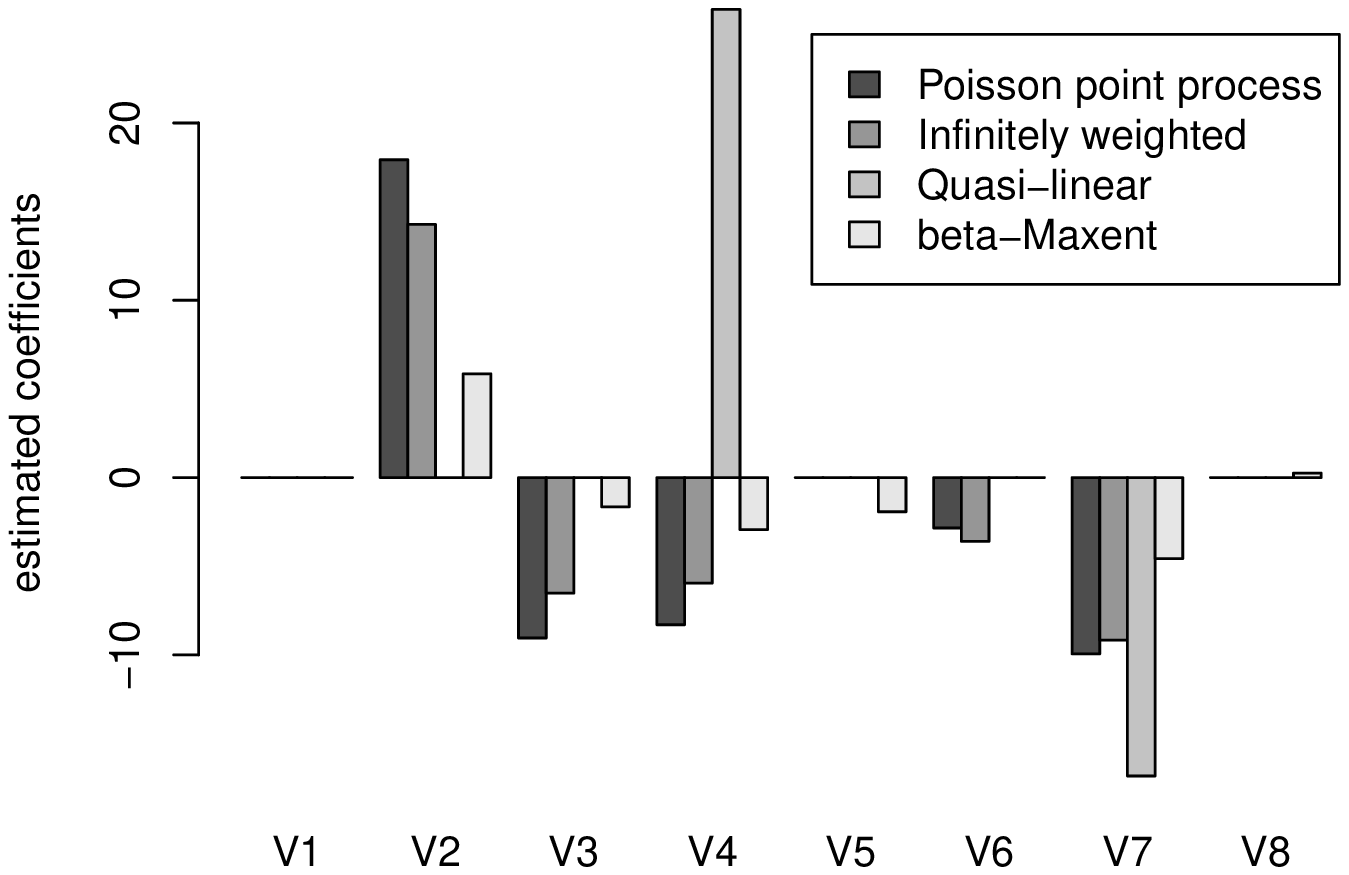}
\caption{Estimated coefficients for Poisson point process, Infinitely weighted logistic regresson, Quasi-linear Poisson regression and $\beta$-Maxent based on V1=mean annual temperature, V2=total annual precipitation, V3=precipitation of wettest quarter, V4=precipitation of driest quarter, V5=max temperature of warmest month, V6=min temperature of coldest month, V7=temperature annual range, and V8=mean temperature of wettest quarter. }\label{fig2}
\end{figure}

Figure \ref{fig2} illustrates the result of estimated coefficients for all methods. As expected results of Poisson point process and infinitely weighted logistic regression resemble each other. On the other hand results of quasi-linear Poisson regression and $\beta$-Maxent are quite different. The estimated coefficient of $V4$ has a large positive value for a quasi-linear Poisson point process. The estimated coefficients for $\beta$-Maxent take relatively small values showing avoidance of overfitting. The values of AUC are also calculated using background locations as pseudo-absence locations, resulting in around 0.77 for all methods.

\section{Information divergence}\label{ID}

We consider the information divergence for Poisson point processes.
Let $p(\xi)$ and $q(\xi)$ be density functions of two Poisson point processes,
where $\xi=(n,\{s_1,...,s_n\})$ is a realization with the number $n$ and the set $\{s_1,...,s_n\}$ of points.
From the discussion above, the density functions are written as
\begin{eqnarray}\label{pq}
p(\xi) =
\exp\{-\Lambda \}\prod_{i=1}^n {\lambda(s_i)}, \ \ \ q(\xi) =
\exp\{-H \}\prod_{i=1}^n {\eta(s_i)},
\end{eqnarray}
in which $p(\xi)$ and $\lambda(s)$ have a one-to-one correspondence, and
 $q(\xi)$ and $\eta(s)$ have also the same correspondence.
The Kullback-Leibler (KL) divergence between $p$ and $q$ is defined by the difference between the cross entropy
and the diagonal entropy as ${\cal D}_{\rm KL}(p,q)={\cal H}(p,q)-{\cal H}(p,p)$, where the cross entropy is defined by
$${\cal H}(p,q)=-\mathbb E_{p} [\log {q(\Xi)}],$$
where $\mathbb E_p$ is the expectation with the density function $p(\xi)$.
Thus, it is written as  
\begin{eqnarray}\label{kl}
{\cal D}_{\rm KL}(p,q)= \int_A\Big\{ \lambda(s) \log \frac{\lambda(s)}{\eta(s)}-\lambda(s)+\eta(s)\Big\}{\rm d}s
\end{eqnarray}
since the cross entropy is written as
\begin{eqnarray}\label{eq42}
{\cal H}(p,q)
&=&-\sum_{n=0}^\infty \frac{\Lambda^n}{n!}e^{-\Lambda}
\int_A\cdots\int_A  \log \{e^{-H}\prod_{j=1}^n {\eta(s_j)}\}\prod_{j=1}^n\frac{\lambda(s_j)}{\Lambda}{\rm d}s_j
\nonumber\\[2mm]
&=&\sum_{n=0}^\infty \frac{\Lambda^n}{n!}e^{-\Lambda}
\Big[H - \frac{n}{\Lambda}
 \int_A {\lambda(s)}\log{\eta(s)}{\rm d}s\Big]
 \nonumber\\[2mm]
&=& \int_A\{ -\lambda(s) \log {\eta(s)}+\eta(s)\}{\rm d}s,
\end{eqnarray}
where $H=\int_A \eta(s)ds$.
We note from \eqref{kl} that ${\cal D}_{\rm KL}(p,q)$ coincides with the extended KL divergence between intensity functions
$\lambda$ and $\eta$, say $D_{\rm eKL}(\lambda,\eta)$.
Here, the term $-\lambda(s)+\eta(s)$ in the integrand of \eqref{kl} should be added
to the standard form since both $\lambda(s)$ and $\eta(s)$ in general do not have total mass one. 
If $q$ is assumed a parametric model as
\begin{eqnarray}
p(\xi,\theta) =
\exp\{-\Lambda(\theta) \}\prod_{i=1}^n {\lambda(s_i,\theta)},
\end{eqnarray}
then the  empirical counterpart of the cross entropy ${\cal H}(p,p(\cdot,\theta))$ leads to a loss function
\begin{eqnarray}\label{neg}
L_{\rm KL}(\theta)=-\sum_{i=1}^n \log \lambda(s_i,\theta)+\Lambda(\theta)
\end{eqnarray}
replacing the expectation $\mathbb E_p$ to the empirical expectation for a presence dataset $\{s_1,...,s_n\}$. 
It is noted that \eqref{neg} is nothing but the negative log-likelihood as seen in \eqref{ell}.
Note that, if the realization $(n,\{s_1,...,s_n\})$ is generated from $p(\xi)$, then
we conclude $\mathbb E_p[L_{\rm KL}(\theta)]={\cal H}(p,p(\cdot,\theta))$ due to
a basic formula of a random sum in the Poisson point process in the same way as (\ref{eq42}), 
\begin{eqnarray}
\mathbb E_p\Big[\sum_{i=1}^N \log \lambda(S_i,\theta)\Big]
=\int_A \lambda(s)\log \lambda(s,\theta){\rm d}s
\end{eqnarray}
for $\Xi=(N,\{S_1,...,S_N\})$.
This guarantees the consistency of the maximum likelihood estimator (MLE) for $\theta$
under an assumption where the true density function $p(\xi)$ is equal to the model
$p(\xi,\theta)$.
Here, we note 
\begin{eqnarray}
\mathbb E_{p(\cdot,\theta)}[\ell(\theta)]-\mathbb E_{p(\cdot,\theta)} [\ell(\tilde\theta) ]
={\cal D}_{\rm KL}(p(\cdot,\theta),p(\cdot,\tilde\theta))
\end{eqnarray}
which is greater than or equal to $0$ for any $\tilde\theta$ of $\Theta$, and the equality holds if and only if
$\tilde\theta=\theta$.
In general, it is known that the maximum likelihood is equivalent to the minimum KL divergence, see \cite{Akaike1974} for more general discussion. 

We observe an interesting relationship between the pair of density functions $p(\xi)$ and $q(\xi)$ given in \eqref{pq} and the pair of the intensity functions $\lambda(s)$ and $\eta(s)$ such that
$D_{\rm KL}(p,q)=D_{\rm eKL}(\lambda,\eta)$.
Hence we discuss an information divergence class that is defined on the space of intensity functions in place of the space of density functions of Poisson point processes. 
Consider the $\beta$-power divergence defined by the difference between the $\beta$-cross entropy and the $\beta$-diagonal entropy as $D_{\beta}(\lambda,\eta)=H_{\beta}(\lambda,\eta)-H_{\beta}(\lambda,\lambda)$, where
\begin{eqnarray}
H_{\beta}(\lambda,\eta)=-\frac{1}{\beta}\int \lambda(s)\eta(s)^\beta {\rm d}s
+\frac{1}{\beta+1}\int \eta(s)^{\beta+1} {\rm d}s.
\end{eqnarray}
See \cite{Basu1998, Minami2002} for the $\beta$-power divergence, however we apply to the space of intensity functions rather than the space of density functions.
Then, by analogy with the KL divergence, the $\beta$-power loss function based on the presence dataset $\{s_1,...,s_n\}$ is given by
\begin{eqnarray}\label{beta-loss}
L_{\beta}(\theta)=-\frac{1}{\beta}\sum_{i=1}^n  \lambda(s_i,\theta)^\beta+
\frac{1}{\beta+1}\int \lambda(s,\theta)^{\beta+1} {\rm d}s
\end{eqnarray}
and the minimum $\beta$-power divergence estimator is defined by 
$\hat\theta_\beta=\argmin_{\theta\in{\Theta}}L_\beta(\theta)$.
The second term of $L_\beta(\theta)$ is approximated by the Gaussian quadrature similar to the log-likelihood function. 
Assume that  the realization $(n,\{s_1,...,s_n\})$ is generated from $p(\xi)$ with the intensity function $\lambda(s)$.
Then,
\begin{eqnarray}
\mathbb E_p[L_{\beta}(\theta)]=H_\beta(\lambda,\lambda(\cdot,\theta)),
\end{eqnarray}
which also guarantees the consistency of $\hat\theta_\beta$ for $\theta$ 
under an assumption where the true density function $p(\xi)$ is equal to $p(\xi,\theta)$ with the intensity function $\lambda(s,\theta)$.
The approximate estimating equation by the quadrature is given by
\begin{eqnarray}\label{beta-eq}
{\cal E}_\beta(\theta):= \sum_{i=1}^m w_i \lambda(s_i,\theta)^\beta \{
 \zeta_i- \lambda(  s_i , \theta)\}
  \frac{\partial}{\partial \theta}\log  \lambda(  s_i , \theta)
=0,
\end{eqnarray}
where $w_i$ is the quadrature weight.
Thus, the estimating equation is the weighted likelihood equation for \eqref{like} with the weight function
$\lambda(s_i,\theta)^\beta$.
If we take a limit of $\beta$ to $0$, then $D_\beta(\lambda,\eta)$, $L_\beta(\theta)$ and ${\cal E}_\beta(\theta)$ are equal to $D_{\rm eKL}(\lambda,\eta)$, $-\ell(\theta)$ and ${\cal E}(\theta)$,
respectively, that is, the minimum $\beta$-power divergence is reduced to the maximum likelihood.

We next consider
 the $\gamma$-power divergence defined by the difference between the $\gamma$-cross entropy and the $\gamma$-diagonal entropy as $D_{\gamma}(\lambda,\eta)=H_{\gamma}(\lambda,\eta)-H_{\gamma}(\lambda,\lambda)$, where
\begin{eqnarray}
H_{\gamma}(\lambda,\eta)=-\frac{1}{\gamma}
\frac{\int \lambda(s)\eta(s)^\gamma {\rm d}s}
{\large\{\int \eta(s)^{\gamma+1} {\rm d}s\large\}^{\frac{\gamma}{\gamma+1}}}.
\end{eqnarray}
Similarly, the $\gamma$-power loss function based on the presence dataset $\{s_1,...,s_n\}$ is given by
\begin{eqnarray}
L_{\gamma}(\theta)=-\frac{1}{\gamma}\frac{\sum_{i=1}^n  \lambda(s_i,\theta)^\gamma} {\large\{\int \lambda(s,\theta)^{\gamma+1} {\rm d}s\large\}^{\frac{\gamma}{\gamma+1}}}
\end{eqnarray}
and the minimum $\gamma$-power divergence estimator is defined by 
$\hat\theta_\gamma =\argmin_{\theta\in{\Theta}}L_\gamma(\theta)$.
See \cite{Fujisawa2008}. 
We note that the definition of $H_\gamma(\lambda,\eta)$ is given by the standard form other than the log form, see \cite{Eguchi2022} for the detailed discussion. 
An argument similar to that above yields $\mathbb E_p[L_{\gamma}(\theta)]=H_\gamma(\lambda,\lambda(\cdot,\theta)),$
which also guarantees the consistency of $\hat\theta_\gamma$ for $\theta$. 
The approximate estimating equation is given by
\begin{eqnarray}
{\cal E}_\gamma(\theta):= \sum_{i=1}^m w_i  \{
 \zeta_i u_\gamma(s_i,\theta)-v_\gamma(s_i,\theta) \}
  \frac{\partial}{\partial \theta}\log  \lambda(  s_i , \theta)
=0,
\end{eqnarray}
where
\begin{eqnarray}
u_\gamma(s ,\theta)=\frac{  \lambda(  s  , \theta)^\gamma}{ \sum_{i=1}^n \lambda(  s_i  , \theta)^\gamma}, \  v_\gamma(s ,\theta)=\frac{  \lambda(  s  , \theta)^{\gamma+1}}{ \sum_{i=1}^m w_i \lambda(  s_i  , \theta)^{\gamma+1}}.
\end{eqnarray}
The estimating equation has a property different from that of the estimating equation \eqref{beta-eq}.
We observe
\begin{eqnarray}
\lim_{\gamma\rightarrow0}D_\gamma(\lambda,\eta)=\int_A
\lambda(s)\Big\{\log\frac{\lambda(s)}{\Lambda}-\log\frac{\eta(s)}{H}\Big\}{\rm d}s,
\end{eqnarray}
which is the KL divergence $D_{\rm KL}(\lambda/\Lambda,\eta/H)$.
Thus, the $\gamma$-power loss function is reduced to the negative log-likelihood for the random sample model $\tilde{\cal M}=\{\lambda(s,\theta)/\Lambda(\theta):\theta\in\Theta\}$,
\begin{eqnarray}
\tilde\ell(\theta)=
\sum_{i=1}^n \log \lambda(s_i,\theta)-\log \Lambda(\theta).
\end{eqnarray}
This property exactly coincides with the Maxent, in which 
the Maxent is equivalent to the MLE for the SDM \eqref{SDM} up to the normalization constant,
see Reneer \& Warton (2013) for the detailed discussion.

We have discussed the $\beta$-power 
on the space of intensity functions, in which  the minimum $\beta$-power divergence estimator is an extension of the ML estimator for the Poisson point process model  \eqref{SDM}. 
The estimation methods can be viewed as a weighted likelihood method with the equation as in \eqref{beta-eq}.  
We consider an extension of the $\beta$-power divergence to $U$-divergence.
Let $U(s)$ be a convex function defined on $\mathbb R$. 
Thus, the $U$-divergence  is defined by 
$ D_U(\lambda,\eta)= H_U(\lambda,\eta)- H_U(\lambda,\lambda)$, where
\begin{eqnarray}
  H_U(\lambda,\eta) = \int_{A}\{ U(\xi(\eta(s)))-\xi(\eta(s))\lambda(s) \}{\rm d}s
\end{eqnarray}
for intensity functions $\lambda(s)$ and $\eta(s)$ defined on a study area $A$, 
where $\xi$ is the inverse function of the derivative of $U$, see \cite{Eguchi2022} for the general discussion of $U$-divergence.
Note that $D_U(\lambda,\eta)\geq 0$ 
due to the convexity for $U$  and the equality holds if and only if $\lambda=\eta$
on $A$.
This is because, for any scalars $\tilde\lambda$ and $\tilde\eta$
\begin{eqnarray}\label{lhs}
 U(\xi(\tilde\eta))-U(\xi(\tilde\lambda))-\tilde\lambda\{\xi(\tilde\eta)-\xi(\tilde\lambda)\}\geq0 
\end{eqnarray}
and $D_U(\lambda,\eta)$ is equal to the integral of the left-hand-side of \eqref{lhs}
substituted $\tilde\lambda$ and $\tilde\eta$ into $\lambda(s)$ and $\eta(s)$.
If  $U(t)=\exp(t)$, then the  $D_U(\lambda,\eta)$ is reduced to the extended KL divergence $D_{\rm eKL}(\lambda,\eta) $ in (\ref{kl});
if  
\begin{eqnarray}\label{U-beta}
U(t)=\frac{1}{\beta+1}(1+\beta t)^{\frac{\beta+1}{\beta}},
\end{eqnarray} 
then  $D_U(\lambda,\eta)$ is reduced to the $\beta$-power divergence.
The $U$-loss function for a given location data $\{s_1,...,s_n\}$ is introduced by 
\begin{eqnarray}
  L_U(\theta)=- \sum_{i=1}^n\xi(\lambda(s_i,\theta)) + \int_{A}U(\xi( \lambda(s,\theta))){\rm d}s
\end{eqnarray}
and the minimum $U$-divergence estimator $\hat\theta_U$ is defined by the minimizer
of $L_U(\theta)$ with $\theta$.
The estimating function for $\hat\theta_U$ is given by
\begin{eqnarray}
{\cal E}_U(\theta) = - \sum_{i=1}^n \xi^\prime(\lambda(s_i,\theta))\frac{\partial}{\partial \theta}\lambda(s_i,\theta) + \int_{A} \lambda(s,\theta)\xi^\prime(\lambda(s,\theta))\frac{\partial}{\partial \theta}\lambda(s,\theta) {\rm d}s.
\end{eqnarray}
This estimating function is unbiased, that is, 
$\mathbb E_{\theta}[{\cal E}_U(\theta)] =0$, where $\mathbb E_\theta$ is the expectation with  the model intensity function $\lambda(s,\theta)$.
Thus, the quadrature approximation leads to the estimating equation
\begin{eqnarray}
{\cal E}_U(\theta) = \sum_{i=1}^m w_i \omega(s_i,\theta)  \{ \zeta_i- \lambda(  s_i , \theta)\} \frac{\partial}{\partial \theta}\log  \lambda(s_i , \theta)=0,
\end{eqnarray}
where $\omega(s,\theta)=\xi'(\lambda(s_i,\theta))\lambda(s,\theta)$.
If $U$ is adopted as \eqref{U-beta}, then $U$-loss function is nothing but 
the $\beta$-power loss function \eqref{beta-loss} and the estimating function \eqref{beta-eq}.
Thus,  the corresponding weight function is $\omega(s,\theta)=\lambda(s,\theta)^\beta$.
For example, if a log-linear model is assumed as $\lambda(s,\theta)=\exp\{\theta^\top
x(s)\}$ with a feature vector $x(s)$, then the weight function is not a bounded 
function of $x(s)$. 
This shows the minimum $\beta$-power divergence method is concerned about an unpreferable behavior. 
For this issue, we employ a cumulative distribution function on a nonegative random variable.
Assume that the derivative of the generator function is given by
\begin{eqnarray}
  \xi(t) = \int_0^t\frac{F(\tau u)}{u}{\rm d}u,
\end{eqnarray}
where $\tau>0$ is a constant and $F(\cdot)$ is a cumulative distribution function (cdf).
Then the estimating equation has a cdf-weighted form as follows:
\begin{eqnarray}
{\cal E}_U(\theta) = \sum_{i=1}^m w_i F( \tau\lambda(s_i,\theta) ) \{ \zeta_i- \lambda(  s_i , \theta)\} \frac{\partial}{\partial \theta}\log  \lambda(s_i , \theta)=0.
\end{eqnarray}

\section{Integrated SDMs}

We discuss an estimation method for a model integrating SDMs. 
It frequently appears in ecological studies that  composite datasets for a target species are observed by
different occasions and mechanisms. 
The integrated model for combining SDMs to such composite datasets
is discussed in the formulation of Poisson point process, which helps modeling jointly these datasets under a reasonable assumption. 
We consider a statistical method for predicting the presence of the species via coupling 
different estimating methods for models based on these datasets.   
The key is to estimate  the shared parameter combining SDMs. 
We discuss a class of estimation methods for selecting an adapted estimation of
the shared parameter in the  integrated SDM.

Consider a typical application for integrating a presence-background (PB) model and a site-occupancy (SO) model, cf. \cite{Koshkina2017} for detailed discussion. 
We suppose that there is a Poisson point process with an intensity function modeled as ${\cal M}_0=\{\lambda_0(s,\beta):\beta\in {\cal B}\}$ with an unknown parameter $\beta$ of the space $\cal B$.
Thus, the intensity function $\lambda_0(s,\beta)$ depends on the site $s$, in which
a log-linear model is commonly assumed as ${\exp\{\beta^\top x(s)\}}$ with
a covariate vector $x(s)$ composed of  environmental variables interacting the habitation.
In the PB model the observation is based on opportunistic sampling.
Hence the detection probability
is also depending on $s$, in which the probability is frequently assumed  to be in a logistic model 
\begin{eqnarray}\label{bias}
  p(s,\alpha)=\frac{\exp\{\alpha^\top v(s)}\}{1+\exp\{\alpha^\top v(s)}\}
\end{eqnarray}
with a covariate vector $v(s)$ composed of  variables  associated with the accessibility to site $s$.
Thus,  the PB model is introduced by  a thinned Poison point process
\begin{eqnarray}\label{pb}
p(n,\{s_1,...,s_n\},\alpha,\beta):=\exp\{- \Lambda(A,\alpha,\beta)\}\prod_{i=1}^n  p(s_i,\alpha)\lambda_0(s_i,\beta)
\end{eqnarray}
with $\Lambda(A,\alpha,\beta)=\int_A p(s,\alpha)\lambda_0(s,\beta) ds$.

In the  SO model the observation is conducted by an experimental design that
is planed in repeated surveys across the predetermined sites and occasions.
The study area is divided into  non-overlapping $K$ regions $C_1,..., C_K$
with $T$ time intervals.
Then, the study is summarized as $Y=\{Y_{ij}:1\leq i\leq K,1\leq j\leq T\}$, where $y_{ij}=1$ if the target species is detected at the $i$ site during survey $j$ and $y_{ij}=0$ otherwise.
Under the assumption for the independence over the sites and intervals, the probability
distribution of the random matrix $Y$ is written as
\begin{eqnarray}\label{so}
f(y,\beta,\tau)=\prod_{i=1}^K\Big[ \Psi_i
  \prod_{j=1}^T p(y_{ij}\mid  z_{ij},\tau)\Big]^{1-S_{i}}
\Big[ \Psi_i  \prod_{j=1}^T p(0\mid  z_{ij},\tau)+1-\Psi_i\Big]^{S_{i}},
\end{eqnarray}
where $y=(y_{ij})_{1\leq i\leq K,1\leq j\leq T}$ and 
$S_{i} =1$ if $\sum_{j=1}^{T}y_{ij}=0$ and $0$ otherwise.
Here $\Psi_i$ is the probability for the species to occupy at region $C_i$ that is
given by 
\begin{eqnarray}
\Psi_i={\rm P}(N(C_i)>0)=1-\exp\bigg[-\int_{C_i}\lambda_0(s,\beta){\rm d}s\bigg].
\end{eqnarray}
due to the basic assumption of the Poisson point process.
The probability that the species is detected in $C_i$ on the $j$-th survey is
typically modeled as
\begin{eqnarray}
p(y_{ij}\mid  z_{ij},\tau)=\frac{\exp(y_{ij}\tau^\top z_{ij})}{1+\exp(\tau^\top z_{ij})}.
\end{eqnarray}
with a covariate vector $z_{ij}$ related to the detection for the species.

For a given set of the location data $\{s_1,...,s_n\}$   and the matrix data $y=\{y_{ij}\}$
the maximum likelihood (ML) is the standard method integrating the PB model \eqref{pb} and the SO model \eqref{so}.
The integrated log-likelihood function is given by $\ell(\theta)_{\rm I}=\ell_{\rm PB}(\beta,\alpha)+\ell_{\rm SO}(\beta,\tau)$, where $\theta=(\beta,\alpha,\tau)$, 
\begin{eqnarray}
\ell_{\rm PB}(\beta,\alpha)= \log p(n,\{s_1,...,s_n\},\alpha,\beta),   \ 
\ell_{\rm SO}(\beta,\tau)=\log f(y,\beta,\tau).
\end{eqnarray}
The ML estimator $\hat\theta_{\rm I}=(\hat\beta_{\rm I},\hat\alpha_{\rm I},\hat\tau_{\rm I})$ for $\theta$ is defined by maximization of
$\ell_{\rm I}(\theta)$ with respect to $\theta$.
The estimating equation is given by ${\cal E}_{\rm I}(\theta)={\cal E}_{\rm PB}(\beta,\alpha)+{\cal E}_{\rm SO}(\beta,\tau) =0$, where
\begin{eqnarray}\label{PBeq}
{\cal E}_{\rm PB}(\beta,\alpha)=\nabla_{(\beta , \alpha )}\log p(n,\{s_1,...,s_n\},\alpha,\beta)
\end{eqnarray}
and
\begin{eqnarray}\label{SOeq}
{\cal E}_{\rm SO}(\beta,\tau)=\nabla_{
(\beta , \tau)}\log f(y,\beta,\tau),
\end{eqnarray}
where $\nabla_\theta$ denotes  gradient vector with $\theta$.
In the compound parameter $\theta$,  $\beta$ is the shared parameter  that simultaneously defines the PB and SO models, whereas $\alpha$ and $\tau$ are parameters separately defining the PB and SO models, respectively.
In effect, we can separately get the ML estimators $(\hat\beta_{\rm PB},\hat\alpha_{\rm PB})$ and $(\hat\beta_{\rm SO},\hat\tau_{\rm SO})$ solving the equations
${\cal E}_{\rm PB}(\beta,\alpha)=0$ and ${\cal E}_{\rm SO}(\beta,\tau)=0$, respectively.
Both of estimators $\hat\beta_{\rm PB}$ and $\hat\beta_{\rm SO}$ are asymptotically consistent for $\beta$.
However, the integrated log-likelihood function has more information about $\beta$
under the assumption of the PB and SO models, and hence the integrated ML estimator $\hat\beta_{\rm I}$ is more efficient than either of $\hat\beta_{\rm PB}$ and $\hat\beta_{\rm SO}$.  The Fisher information matrices ${\cal I}_{\rm I}(\beta)$ for $\beta$ possessed in $\ell_{\rm I}(\beta,\alpha,\tau)$
is the sum of the Fisher information matrices ${\cal I}_{\rm PB}(\beta)$ and ${\cal I}_{\rm SO}(\beta)$ possessed in $\ell_{\rm PB}(\beta,\alpha)$ and  $\ell_{\rm SO}(\beta,\tau)$, where
\begin{eqnarray}
{\cal I}_{\rm PB}(\beta)=- \mathbb E\Big\{
\frac{\partial^2}{\partial\beta\partial\beta^\top}\ell_{\rm PB}(\beta,\alpha)\Big\}, \ \ 
{\cal I}_{\rm SO}(\beta)=- \mathbb E\Big\{
\frac{\partial^2}{\partial\beta\partial\beta^\top}\ell_{\rm SO}(\beta,\tau)\Big\}.
\end{eqnarray}
The asymptotic arguments yield the asymptotic normal properties:
$ 
\sigma_{m,T}({\hat\beta}_{\rm I}(\beta)-\beta)\sim N(0,{\cal I}_{\rm I}(\beta)^{-1})$, 
$\sigma_n({\hat\beta}_{\rm PB}(\beta)-\beta)\sim N(0,{\cal I}_{\rm PB}(\beta)^{-1}),$
and
$
\sigma_{T}({\hat\beta}_{\rm SO}(\beta)-\beta)\sim N(0,{\cal I}_{\rm SO}(\beta)^{-1})
$ 
as $m$ and $T$ go to $\infty$.
See \cite{Rathbun1994} for detailed discussion for the asymptotic properties under the spatial  Poisson point processes.
In accordance, the integrated ML estimator $\hat{\beta}_{\rm I}$
improves the performance of either of the ML estimators $\hat{\beta}_{\rm PB}$ or
$\hat{\beta}_{\rm SO}$.

We discuss more practical situation for the shared parameter $\beta$
that simultaneously defines the PB and SO model.
The qualities of the observation applied to two models are contrast, that is, the observation mechanism for  PB data is opportunistic sampling based on basically 
no predetermined design for the survey including observations by volunteers, whereas the SO sampling is conducted by an organized plan for the survey with predetermined regions and duration times.    
Hence, the PB data may include undetectable outliers due to departure from the PB model however the PB model  introduces the detection probability $p(s,\alpha)$ 
for an observer with the covariate $x(s)$ at  site $s$ in \eqref{bias}.
So, we consider a robust estimating method for the PB model whereas the ML estimator for SO model is fixed as  \eqref{pb}. 
The estimating equation for $\theta=(\beta,\alpha,\tau)$ is proposed as 
\begin{align}\label{F}
{\cal E}_{\rm I}(\theta,F)= \sum_{i=1}^m F(\lambda(s_i,\alpha,\beta)){\cal E}_{\rm PB}(s_i,\zeta_i,w_i,\alpha,\beta)
+
\sum_{i=1}^K\sum_{j=1}^T {\cal E}_{\rm SO}(y_{ij},\beta,\tau)=0,
\end{align}
where ${\cal E}_{\rm PB}(s_i,\zeta_i,w_i,\alpha,\beta)$ is the $i$-th element of ${\cal E}_{\rm PB}(\beta,\alpha)$ and ${\cal E}_{\rm SO}(y_{ij},\beta,\tau)$ is the $ij$-th element of ${\cal E}_{\rm SO}(\beta,\tau)$
%
and $F$ is a cdf  for robust estimation as discussed in Section \ref{ID}.
Note that this  weighting between likelihood equations  \eqref{PBeq} and \eqref{SOeq} is asymmetric such that 
 the weighting of $F(\lambda)$ is conducted only for \eqref{PBeq}.
Thus, the integrated estimating equation \eqref{F} gives
a robust and efficient estimator in the situation discussed above.

We next introduce another situation for integrating a PB model and a distance sampling (DS) model, see \cite{Farr2021}. 
Suppose that there is a Poisson point process with an intensity function $\lambda_0(s,\beta)$.
Two observation mechanisms yields two SDMs  on areas $A$ and $B$ by way of
independent spatial Poisson point processes,
in which one is the PB model as discussed in \eqref{pb}; the other is the DS model
described as
\begin{eqnarray}\label{m2}
(N,Y) \sim  q(n,\{t_1,...,t_n\},\beta,\omega):=\exp\{-\Lambda(B,\beta,\omega)\}\prod_{i=1}^n \pi(t_i,\omega)\lambda_0(t_i,\beta),
\end{eqnarray}
where $\Lambda(\beta,\omega)=\int_B \pi(t,\omega)\lambda_0(t,\beta) {\rm d}t$ and $\pi(t,\omega)$ is  the average detection probability at site $t$.
Here $\pi(t,\omega)$ is  supposed as $\exp\{-\frac{1}{2}d(t)^2/\sigma^2(t,\omega)\}$,
where $d(t)$ is the distance between the midpoint of site $s$ and the transect line
and $\sigma(t,\omega)$ is the scale parameter of the half-normal distribution
modeled  by the parameter $\omega$ with a log-link function.   
For a given location data $\{t_1,...,t_n\}$ the log-likelihood function
is given as 
$\ell_{DS}(\beta,\omega)=\log q(n,\{t_1,...,t_n\},\beta,\omega)$ with
the estimating function ${\cal E}_{DS}(\beta,\omega)=\nabla_{(\beta,\omega)}\ell_{DS}(\beta,\omega)$.
See \cite{Farr2021} for more details of the integrated model.
We can discuss the robust and efficient combination between ${\cal E}_{PB}(\beta,\alpha)$ and ${\cal E}_{DS}(\beta,\omega)$ on the ground that
the DS sampling is more reliable than the PB sampling.

\section{Concluding remarks}
SDM estimates potential habitat maps for target species, which are useful for conservation management \citep{Villero2017}. However, we have to understand SDM has limitations due to sampling biases, imperfect detection, human impacts, range shifting of the species by climate change and so on. In this situation, SDM database plays an important role to obtain reliable estimation of habitat maps \citep{Frans2022}. Available database includes SeaLifeBase (\url{https://www.sealifebase.ca/}), AquaMaps (\url{https://www.aquamaps.org/}), Open Tree of Life (\url{https://tree.opentreeoflife.org/}), Vertlife (\url{https://vertlife.org/}) and so on. In Japan, Ocean 180 Database (\url{https://ocean180-pj.github.io/data.website/index.html}) managed by a research group in University of the Ryukyus provides a wide range of data sets for bioderversity studies. 

The Ocean 180 Database also plays an important role to promote interdisciplinary collaborations between ecologists, biologists, statisticians, computer scientists, business persons and local municipalities. Some of them are studies on species abundance at large spatial scales \citep{Fukaya2020}, a large-scale colonization pattern of exotic seed plants \citep{Kusumoto2021}, the conservation effectiveness of the Japanese protected areas network \citep{Shiono2021}, geometric framework for multiple macroecological patterns \citep{Takashina2019} and so on. We hope that this paper gives fundamental aspects and recent advances of Poisson point process and also contributes to the interdisciplinary collaborations and researches.

\section*{Acknowledgements}
We would like to thank two referees for careful reading and useful suggestions, which much improve quality of our manuscript.
Part of this work is supported by JSPS KAKENHI No. JP18H03211 and No. JP22K11938.

\section*{Declarations}
The authors declare that they have no conflict of interest.

\begin{thebibliography}{50}
\providecommand{\natexlab}[1]{#1}
\providecommand{\url}[1]{\texttt{#1}}
\providecommand{\urlprefix}{URL }

\bibitem[{Akaike(1973)}]{Akaike1973}
Akaike, H. (1973) Information theory and an extension of the maximum likelihood
  principle.
\newblock \emph{Second International Symposium on Information Theory}, pp.
  267--281.

\bibitem[{Akaike(1974)}]{Akaike1974}
Akaike, H. (1974) A new look at the statistical model identification.
\newblock \emph{IEEE Transactions on Automatic Control}, pp. 716--723.

\bibitem[{Basu \emph{et~al.}(1998)Basu, Harris, Hjort \& Jones}]{Basu1998}
Basu, A., Harris, I.R., Hjort, N. \& Jones, M. (1998) Robust and efficient
  estimation by minimising a density power divergence.
\newblock \emph{Biometrika}, \textbf{85}, 549--559.

\bibitem[{Belbin~L(2013)}]{Belbin2013}
Belbin~L, Daly~J, H.T.H.D.S.J. (2013) A specialist's audit of aggregated
  occurrence records: An 'aggregator's' perspective.
\newblock \emph{Zookeys}, \textbf{305}, 67--76.

\bibitem[{Chao \emph{et~al.}(2005)Chao, Chazdon, Colwell \& Shen}]{Chao2005}
Chao, A., Chazdon, R.L., Colwell, R.K. \& Shen, T.J. (2005) A new statistical
  approach for assessing similarity of species composition with incidence and
  abundance data.
\newblock \emph{Ecology Letters}, \textbf{8}, 148--159.

\bibitem[{Copas(1988)}]{Copas1988}
Copas, J. (1988) Binary Regression Models for Contaminated Data.
\newblock \emph{Journal of the Royal Statistical Society: Series B.},
  \textbf{50}, 225--265.

\bibitem[{Dud{\'i}k \emph{et~al.}(2004)Dud{\'i}k, Phillips \&
  Schapire}]{Dudik2004}
Dud{\'i}k, M., Phillips, S.J. \& Schapire, R.E. (2004) Performance Guarantees
  for Regularized Maximum Entropy Density Estimation.
\newblock  \emph{Learning Theory} (eds. J.~Shawe-Taylor \& Y.~Singer), pp.
  472--486. Springer Berlin Heidelberg, Berlin, Heidelberg.

\bibitem[{Dud{\'i}k \emph{et~al.}(2005)Dud{\'i}k, Schapire \&
  Phillips}]{Dudik2005}
Dud{\'i}k, M., Schapire, R.E. \& Phillips, S.J. (2005) Correcting sample
  selection bias in maximum entropy density estimation.
\newblock \emph{Advances in Neural Information Processing System 18},
  \textbf{18}, 323--330.

\bibitem[{Eguchi \& Komori(2015)}]{Eguchi2015}
Eguchi, S. \& Komori, O. (2015) Path Connectedness on a Space of Probability
  Density Functions.
\newblock  \emph{Geometric Science of Information: Second International
  Conference, GSI 2015} (eds. F.~Nielsen \& F.~Barbaresco), p. 615. Springer
  International Publishing, Cham.

\bibitem[{Eguchi \& Komori(2022)}]{Eguchi2022}
Eguchi, S. \& Komori, O. (2022) \emph{Minimum Divergence Methods in Statistical
  Machine Learning: From an Information Geometric Viewpoint}.
\newblock Springer, Tokyo.

\bibitem[{Elith \emph{et~al.}(2006)Elith, Graham, Anderson, Dud{\'i}k, Ferrier,
  Guisan, Hijmans, Huettmann, Leathwick, Lehmann, Li, Lohmann, Loiselle,
  Manion, Moritz, Nakamura, Nakazawa, Overton, Peterson, Phillips, Richardson,
  Scachetti-Pereira, Schapire, Sober{\'o}n, Williams, Wisz \&
  Zimmermann}]{Elith2006}
Elith, J., Graham, C.H., Anderson, R.P., Dud{\'i}k, M., Ferrier, S., Guisan,
  A., Hijmans, R.J., Huettmann, F., Leathwick, J.R., Lehmann, A., Li, J.,
  Lohmann, L.G., Loiselle, B.A., Manion, G., Moritz, C., Nakamura, M.,
  Nakazawa, Y., Overton, J.M., Peterson, A.T., Phillips, S.J., Richardson, K.,
  Scachetti-Pereira, R., Schapire, R.E., Sober{\'o}n, J., Williams, S., Wisz,
  M.S. \& Zimmermann, N.E. (2006) Novel methods improve prediction of species’
  distributions from occurrence data.
\newblock \emph{Ecography}, \textbf{29}, 129--151.

\bibitem[{Farr \emph{et~al.}(2021)Farr, Green, Holekamp \& Zipkin}]{Farr2021}
Farr, M.T., Green, D.S., Holekamp, K.E. \& Zipkin, E.F. (2021) Integrating
  distance sampling and presence-only data to estimate species abundance.
\newblock \emph{Ecology}, \textbf{102}, e03204.

\bibitem[{Fithian \emph{et~al.}(2015)Fithian, Elith, Hastie \&
  Keith}]{Fithian2015}
Fithian, W., Elith, J., Hastie, T. \& Keith, D.A. (2015) Bias correction in
  species distribution models: pooling survey and collection data for multiple
  species.
\newblock \emph{Methods in Ecology and Evolution}, \textbf{6}, 424--438.

\bibitem[{Fithian \& Hastie(2013)}]{Fithian2013}
Fithian, W. \& Hastie, T. (2013) Finite-sample equivalence in statistical
  models for presence-only data.
\newblock \emph{Annals of Applied Statistics}, \textbf{7}, 1917--1939.

\bibitem[{Frans \emph{et~al.}(2022)Frans, Aug\'e, Fyfe, Zhang, McNally,
  Edelhoff, Balkenhol \& Engler}]{Frans2022}
Frans, V.F., Aug\'e, A.A., Fyfe, J., Zhang, Y., McNally, N., Edelhoff, H.,
  Balkenhol, N. \& Engler, J.O. (2022) Integrated SDM database: Enhancing the
  relevance and utility of species distribution models in conservation
  management.
\newblock \emph{Methods in Ecology and Evolution}, \textbf{13}, 243--261.

\bibitem[{Fujisawa \& Eguchi(2008)}]{Fujisawa2008}
Fujisawa, H. \& Eguchi, S. (2008) Robust parameter estimation with a small bias
  against heavy contamination.
\newblock \emph{Journal of Multivariate Analysis}, \textbf{99}, 2053--2081.

\bibitem[{Fukaya \emph{et~al.}(2020)Fukaya, Kusumoto, Shiono, Fujinuma \&
  Kubota}]{Fukaya2020}
Fukaya, K., Kusumoto, B., Shiono, T., Fujinuma, J. \& Kubota, Y. (2020)
  Integrating multiple sources of ecological data to unveil macroscale species
  abundance.
\newblock \emph{Nature Communications}, \textbf{11}, 1695.

\bibitem[{King \& Zeng(2001)}]{King2001}
King, G. \& Zeng, L. (2001) Logistic regression in rare events data.
\newblock \emph{Political Analysis}, \textbf{9}, 137--163.

\bibitem[{Komori \& Eguchi(2014)}]{Komori2014}
Komori, O. \& Eguchi, S. (2014) Maximum power entropy method for ecological
  data analysis.
\newblock  \emph{Bayesian Inference and Maximum Entropy Methods in Science and
  Engineering (Maxent2014)} (eds. A.~Mohammad-Djafari \& F.~Barbaresco), pp.
  337--344. AIP publishing, New York.

\bibitem[{Komori \& Eguchi(2019)}]{Komori2019}
Komori, O. \& Eguchi, S. (2019) \emph{Statistical Methods for Imbalanced Data
  in Ecological and Biological Studies}.
\newblock Springer, Tokyo.

\bibitem[{Komori \emph{et~al.}(2016)Komori, Eguchi, Ikeda, Okamura, Ichinokawa
  \& Nakayama}]{Komori2016}
Komori, O., Eguchi, S., Ikeda, S., Okamura, H., Ichinokawa, M. \& Nakayama, S.
  (2016) An asymmetric logistic regression model for ecological data.
\newblock \emph{Methods in Ecology and Evolution}, \textbf{7}, 249--260.

\bibitem[{Komori \emph{et~al.}(2020)Komori, Eguchi, Saigusa, Kusumoto \&
  Kubota}]{Komori2020}
Komori, O., Eguchi, S., Saigusa, Y., Kusumoto, B. \& Kubota, Y. (2020) Sampling
  bias correction in species distribution models by quasi-linear Poisson point
  process.
\newblock \emph{Ecological Informatics}, \textbf{55}, 1--11.

\bibitem[{Konishi \& Kitagawa(1996)}]{Konishi1996}
Konishi, S. \& Kitagawa, G. (1996) Generalised information criteria in model
  selection.
\newblock \emph{Biometrika}, \textbf{83}, 875--890.

\bibitem[{Koshkina \emph{et~al.}(2017)Koshkina, Wang, Gordon, Dorazio, White \&
  Stone}]{Koshkina2017}
Koshkina, V., Wang, Y., Gordon, A., Dorazio, R.M., White, M. \& Stone, L.
  (2017) Integrated species distribution models: combining presence-background
  data and site-occupancy data with imperfect detection.
\newblock \emph{Methods in Ecology and Evolution}, \textbf{8}, 420--430.

\bibitem[{Kubota \emph{et~al.}(2015)Kubota, Shiono \& Kusumoto}]{Kubota2015}
Kubota, Y., Shiono, T. \& Kusumoto, B. (2015) Role of climate and geohistorical
  factors in driving plant richness patterns and endemicity on the east Asian
  continental islands.
\newblock \emph{Ecography}, \textbf{38}, 639--648.

\bibitem[{Kusumoto \emph{et~al.}(2021)Kusumoto, Kubota, Shiono \&
  Villalobos}]{Kusumoto2021}
Kusumoto, B., Kubota, Y., Shiono, T. \& Villalobos, F. (2021) Biogeographical
  origin effects on exotic plants colonization in the insular flora of Japan.
\newblock \emph{Biological Invasions}, \textbf{23}, 2973--2984.

\bibitem[{Maalouf \& Siddiqi(2014)}]{Maalouf2014}
Maalouf, M. \& Siddiqi, M. (2014) Weighted logistic regression for large-scale
  imbalanced and rare events data.
\newblock \emph{Knowledge-Based Systems}, \textbf{59}, 142--148.

\bibitem[{Maalouf \& Trafalis(2011)}]{Maalouf2011}
Maalouf, M. \& Trafalis, T.B. (2011) Robust weighted kernel logistic regression
  in imbalanced and rare events data.
\newblock \emph{Computational Statistics and Data Analysis}, \textbf{55},
  168--183.

\bibitem[{Manski \& Lerman(1977)}]{Manski1977}
Manski, C.F. \& Lerman, S.R. (1977) The estimation of choice probabilities from
  choice based samples.
\newblock \emph{Econometrica}, \textbf{45}, 1977--1988.

\bibitem[{McCullagh \& Nelder(1989)}]{McCullagh1989}
McCullagh, P. \& Nelder, J. (1989) \emph{Generalized Linear Models}.
\newblock Chapman \& Hall, New York.

\bibitem[{Mesibov(2013)}]{Mesibov2013}
Mesibov, R. (2013) A specialist’s audit of aggregated occurrence records.
    \newblock \emph{ZooKeys}, \textbf{293}, 11--18.

\bibitem[{Minami \& Eguchi(2002)}]{Minami2002}
Minami, M. \& Eguchi, S. (2002) Robust blind source separation by beta
  divergence.
\newblock \emph{Neural Computation}, \textbf{14}, 1859--1886.

\bibitem[{Murata \emph{et~al.}(2004)Murata, Takenouchi, Kanamori \&
  Eguchi}]{Murata2004}
Murata, N., Takenouchi, T., Kanamori, T. \& Eguchi, S. (2004) Information
  geometry of ${\mathcal{U}}$-boost and Bregman divergence.
\newblock \emph{Neural Computation}, \textbf{16}, 1437--1481.

\bibitem[{Naudts(2011)}]{Naudts2011}
Naudts, J. (2011) \emph{Generalised thermostatistics}.
\newblock Springer Science \& Business Media, Berlin.

\bibitem[{Phillips \& Dud{\'i}k(2008)}]{Phillips2008}
Phillips, S.J. \& Dud{\'i}k, M. (2008) Modeling of species distributions with
  Maxent: new extensions and a comprehensive evaluation.
\newblock \emph{Ecography}, \textbf{31}, 161--175.

\bibitem[{Phillips \emph{et~al.}(2004)Phillips, Dud{\'i}k \&
  Schapire}]{Phillips2004}
Phillips, S.J., Dud{\'i}k, M. \& Schapire, R.E. (2004) A Maximum Entropy
  Approach to Species Distribution Modeling.
\newblock \emph{Proceedings of the 21st International Conference on Machine
  Learning. ACM Press, New York}, pp. 472--486.

\bibitem[{Rathbun \& Cressie(1994)}]{Rathbun1994}
Rathbun, S.L. \& Cressie, N. (1994) Asymptotic Properties of Estimators for the
  Parameters of Spatial Inhomogeneous Poisson Point Processes.
\newblock \emph{Advances in Applied Probability}, \textbf{26}, 122--154.

\bibitem[{Renner \& Warton(2013)}]{Renner2013}
Renner, I.W. \& Warton, D.I. (2013) Equivalence of MAXENT and Poisson point
  process models for species distribution modeling in ecology.
\newblock \emph{Biometrics}, \textbf{69}, 274--281.

\bibitem[{Renner \emph{et~al.}(2015)Renner, Elith, Baddeley, Fithian, Hastie,
  Phillips, Popovic \& I.Warton}]{Renner2015}
Renner, I., Elith, J., Baddeley, A., Fithian, W., Hastie, T., Phillips, S.J.,
  Popovic, G. \& I.Warton, D. (2015) Point process models for presence-only
  analysis.
\newblock \emph{Methods in Ecology and Evolution}, \textbf{6}, 366--379.

\bibitem[{Royle \& Dorazio(2008)}]{Royle2008}
Royle, J.A. \& Dorazio, R.M. (2008) \emph{Hierachical Modeling and Inference in
  Ecology: The Analysis of Data from Populations, Metapopulations and
  Communities}.
\newblock Academic Press, London.

\bibitem[{Shiono \emph{et~al.}(2021)Shiono, Kubota \& Kusumoto}]{Shiono2021}
Shiono, T., Kubota, Y. \& Kusumoto, B. (2021) Area-based conservation planning
  in Japan: The importance of OECMs in the post-2020 Global Biodiversity
  Framework.
\newblock \emph{Global Ecology and Conservation}, \textbf{30}, e01783.

\bibitem[{Streit(2010)}]{Streit2010}
Streit, R.L. (2010) \emph{Poisson Point Processes: Imaging, Tracking, and
  Sensing}.
\newblock Springer, New York.

\bibitem[{Takashina \emph{et~al.}(2019)Takashina, Kusumoto, Kubota \&
  Economo}]{Takashina2019}
Takashina, N., Kusumoto, B., Kubota, Y. \& Economo, E.P. (2019) A geometric
  approach to scaling individual distributions to macroecological patterns.
\newblock \emph{Journal of Theoretical Biology}, \textbf{461}, 170--188.

\bibitem[{Villero \emph{et~al.}(2017)Villero, Pla, Camps, Ruiz-Olmo \&
  Brontons}]{Villero2017}
Villero, D., Pla, M., Camps, D., Ruiz-Olmo, J. \& Brontons, L. (2017)
  Integrating species distribution modelling into decision-making to inform
  conservation actions.
\newblock \emph{Biodiversity and Conservation}, \textbf{26}, 251--271.

\bibitem[{Wainer \emph{et~al.}(2007)Wainer, Bradlow \& Wang}]{Wainer2007}
Wainer, H., Bradlow, E.T. \& Wang, X. (2007) \emph{Testlet Response Theory and
  Its Applications}.
\newblock Cambridge University Press, New York.

\bibitem[{Warton(2015)}]{Warton2015}
Warton, D.I. (2015) New opportunities at the interface between ecology and
  statistics.
\newblock \emph{Methods in Ecology and Evolution}, \textbf{6}, 363--365.

\bibitem[{Warton \& McGeoch(2017)}]{Warton2017}
Warton, D.I. \& McGeoch, M.A. (2017) Technical advances at the interface
  between ecology and statistics: improving the biodiversity knowledge
  generation workflow.
\newblock \emph{Methods in Ecology and Evolution}, \textbf{8}, 396--397.

\bibitem[{Warton \& Shepherd(2010)}]{Warton2010}
Warton, D.I. \& Shepherd, L.C. (2010) Poisson point process models solve the"
  pseudo-absence problem" for presence-only data in ecology.
\newblock \emph{The Annals of Applied Statistics The Annals of Applied
  Statistics}, \textbf{4}, 1383--1402.

\bibitem[{Yee(2015)}]{Thomas2015}
Yee, T.W. (2015) \emph{Vector Generalized Linear and Additive Models}.
\newblock Springer, New York.

\bibitem[{Yee \& Mitchell(1991)}]{Thomas1991}
Yee, T.W. \& Mitchell, N.D. (1991) Generalized additive models in plant
  ecology.
\newblock \emph{Journal of Vegetation Science}, \textbf{2}, 587--602.

\end{thebibliography}
\end{document}